# Experimental measurement of plasmonic nanostructures embedded in silicon waveguide gaps

**Alba Espinosa-Soria , Amadeu Griol and Alejandro Martínez**[*]

*Nanophotonics Technology Center, Universitat Politècnica de València, Camino de Vera s/n, Valencia, 46022, Spain.*
[*]*amartinez@ntc.upv.es*

**Abstract:** In this work, we report numerical simulations and experiments of the optical response of a gold nanostrip embedded in a silicon strip waveguide gap at telecom wavelengths. We show that the spectral features observed in transmission and reflection when the metallic nanostructure is inserted in the gap are extremely different to those observed in free-space excitation. First, we find that interference between the guided field and the electric dipolar resonance of the metallic nanostructure results in high-contrast (> 10) spectral features showing an asymmetric Fano spectral profile. Secondly, we reveal a crossing in the transmission and reflection responses close to the nanostructure resonance wavelength as a key feature of our system. This approach, which can be realized using standard semiconductor nanofabrication tools, could lead to fully exploit the extreme properties of subwavelength metallic nanostructures in an on-chip configuration, with special relevance in fields such as biosensing or optical switching.

**References and links**

1. A. V. Zayats and I. I. Smolyaninov, "Near-field photonics: surface plasmon polaritons and localised surface plasmons," J. Opt. A, Pure Appl. Opt. **5**(4), S16–S50 (2003).
2. D. K. Gramotnev and S. I. Bozhelvonyi, "Plasmonics beyond the diffraction limit," Nat. Photonics **4**(2), 83–91 (2010).
3. J. A. Schuller, E. S. Barnard, W. Cai, Y. C. Jun, J. S. White, and M. L. Brongersma, "Plasmonics for extreme light concentration and manipulation," Nat. Materials **9**, 193-204 (2010).
4. J. N. Anker, W. P. Hall, O. Lyandres, N. C. Shah, J. Zhao and R. P. Van Duyne, "Biosensing with plasmonic nanosensors," Nat. Materials **7**, 442–453 (2008).
5. M Kauranen, A.V. Zayats, "Nonlinear plasmonics," Nat. Photonics **6**(11), 737-748 (2010).
6. L. Novotny and N. F. van Hulst, "Antennas for light," Nat. Photonics **5**, 83–90 (2011).
7. M. Husnik, M. W. Klein, N. Feth, M. König, J. Niegemann, K. Busch, S. Linden, and M. Wegener, "Absolute Extinction Cross Section of Individual Magnetic Split-Ring Resonators," Nat. Photonics **2**(10), 614–617 (2008).
8. P.J. Rodríguez-Cantó, M. Martínez-Marco, F. J. Rodríguez-Fortuño, B. Tomás-Navarro, R. Ortuño, S. Peransí-Llopis, and A. Martínez, "Demonstration of near infrared gas sensing using gold nanodisks on functionalized silicon," Opt. Express **19**(8), 7664-7672 (2011).
9. P. Fan, Z. Yu, S. Fan and M. L. Brongersma, "Optical Fano resonance of an individual semiconductor nanostructure," Nat. Materials **13**, 471–475 (2014).
10. H. Aouani, M. Rahmani, M. Navarro-Cía, and S. A. Maier, "Third-harmonic-upconversion enhancement from a single semiconductor nanoparticle coupled to a plasmonic antenna," Nat. Nanotechnology **9**, 290-294 (2014).
11. M. Lipson, "Guiding, modulating, and emitting light on silicon - Challenges and opportunities," J. Lightwave Technol. **23**, 4222-4238 (2005)
12. M. Hochberg and T. Baehr-Jones, "Toward fabless silicon photonics," Nat. Photonics 4, 492-494 (2010).
13. A. Rickman, "The commercialization of silicon photonics," Nat. Photonics **8**, 579-582 (2014).
14. N. Daldosso, M. Melchiorri, F. Riboli, F. Sbrana, L. Pavesi, G. Pucker, C. Kompocholis, M. Crivellari, P. Belluti, and A. Lui, "Fabrication and optical characterization of thin two-dimensional Si3N4 waveguides," Mater. Sci. Semicond. Process. **7**, 453–458 (2004).
15. S. Romero-García, F. Merget, F. Zhong, H. Finkelstein, and J. Witzens, "Silicon nitride CMOS-compatible platform for integrated photonics applications at visible wavelengths," Opt. Express **21**, 14036-14046 (2013)

16. I. Alepuz-Benache, C. García-Meca, F. J. Rodríguez-Fortuño, R. Ortuño, M. Lorente-Crespo, A. Griol, and A. Martínez, "Strong magnetic resonance of coupled aluminum nanodisks on top of a silicon waveguide," Proc. SPIE **8424**, 84242J (2012).
17. F. B. Arango, A. Kwadrin, and A. F. Koenderink, "Plasmonic antennas hybridized with dielectric waveguides," ACS Nano **6**, 10156 (2012).
18. M. Février, P. Gogol, A. Aassime, R. Mégy, C. Delacour, A. Chelnokov, A. Apuzzo, S. Blaize, J-M. Lourtioz, and B. Dagens, "Giant coupling effect between metal nanoparticle chain and optical waveguide," Nano Lett. **12**, 1032 (2012).
19. M. Chamanzar, Z. Xia, S. Yegnanarayanan, and A. Adibi, "Hybrid integrated plasmonic-photonic waveguides for on-chip localized surface plasmon resonance (LSPR) sensing and spectroscopy," Opt. Express **21**, 32086 (2013).
20. F. Peyskens, A. Z. Subramanian, P. Neutens, A. Dhakal, P. V. Dorpe, N. Le Thomas, and R. Baets, "Bright and dark plasmon resonances of nanoplasmonic antennas evanescently coupled with a silicon nitride waveguide," Opt. Express **23**, 3088 (2015).
21. F. Bernal Arango, R. Thijssen, B. Brenny, T. Coenen and A. Femius Koenderink, "Robustness of plasmon phased array nanoantennas to disorder," Sci. Rep. **5**, 10911 (2015).
22. F. Peyskens, A. Dhakal, P. Van Dorpe, N. Le Thomas, and R. Baets, "Surface enhanced Raman spectroscopy using a single mode nanophotonic-plasmonic platform," ACS Photon. **3**, 102 (2016).
23. M. Castro-Lopez, N. de Sousa, A. Garcia-Martin, F. Y. Gardes, and R. Sapienza, "Scattering of a plasmonic nanoantenna embedded in a silicon waveguide.," Opt. Express **23**(22), 28108–18 (2015).
24. M. W. Klein, C. Enkrich, M. Wegener, C. M. Soukoulis, and S. Linden, "Single-slit split-ring resonators at optical frequencies: limits of size scaling," Opt. Lett. **31**, 1259-1261 (2006).
25. N. Verellen, Y. Sonnefraud, H. Sobhan, F. Hao, V. V. Moshchalkov, P. Van Dorpe, P. Nordlander and Stefan A. Maier, "Fano resonances in individual coherent plasmonic nanocavities," Nano Lett. **9**, 1663-1667 (2009).
26. D. Vercruysse, Y. Sonnefraud, N. Verellen, F. B. Fuchs, G. Di Martino, L. Lagae, V. V. Moshchalkov, S. A. Maier, and P. Van Dorpe, "Unidirectional side scattering of light by a single-element nanoantenna," Nano Lett. **13**, 3843-3849 (2013).
27. I. M. Hancu, A. G. Curto, M. Castro-López, M. Kuttge, and N. F. van Hulst, "Multipolar interference for directed light emission," Nano Lett. **14**, 166-171 (2014).
28. A. Espinosa-Soria and A. Martinez, "Transverse spin and spin-orbit coupling in silicon waveguides," http://arxiv.org/abs/1507.04859
29. F. J. Rodríguez-Fortuño, D. Puerto, A. Griol, L. Bellieres, J. Martí, and A. Martínez, "Sorting linearly polarized photons with a single scatterer," Opt. Lett. **39**, 1394-1397 (2014).
30. F. J. Rodríguez-Fortuño, I. Barber-Sanz, D. Puerto, A. Griol, A. Martinez, "Resolving light handedness with an on-chip silicon microdisk", ACS Photonics **1**(9), 762–767 (2014).
31. P. B. Johnson and R. W. Christy, "Optical Constants of the Noble Metals," Phys. Rev. B **6**, 4370 (1972).
32. U. K. Chettiar, R. Fernandez-Garcia, S. A. Maier, and N. Engheta, "Enhancement of radiation from dielectric waveguides using resonant plasmonic coreshells," Opt. Express **20**(14), 16104-16112 (2012).
33. B. Luk'yanchuk, N. I. Zheludev, S. A. Maier, N. J. Halas, P. Nordlander, H. Giessen, and C. T. Chong, "The Fano resonance in plasmonic nanostructures and metamaterials," Nat. Mater. **9**(9), 707–15 (2010).
34. S. Hamed Shams Mousavi, Ali A. Eftekhar, Amir H. Atabaki, and Ali Adibi, "Band-edge bilayer plasmonic nanostructure for surface enhanced Raman spectroscopy," ACS Photon. **2**, 1546-1551 (2015).
35. N. Verellen, F. López-Tejeira, R. Paniagua-Domínguez, D. Vercruysse, D. Denkova, L. Lagae, P. Van Dorpe, V. V. Moshchalkov, and J. a. Sánchez-Gil, "Mode parity-controlled fano- and lorentz-like line shapes arising in plasmonic nanorods," Nano Lett. **14**(5), 2322–2329 (2014).

## 1. Introduction

Metallic nanostructures supporting localized surface plasmons have become key elements for manipulating light fields at the nanoscale [1-3]. Their ability for extreme concentration of the optical field in deep subwavelength volumes, which naturally boost any kind of optical interaction, has been used to implement highly sensitive miniaturized biosensors [4] as well as to demonstrate enhanced nonlinear behavior [5] amongst other applications. Taking profit of their scattering properties, they can also be used efficient transducers between confined and radiated fields, then being termed optical antennas (or nanoantennas) due to their resemblance with their radiofrequency counterparts [6]. Owing to its ultrasmall size, experimental characterization of individual nanostructures supporting plasmonic resonances is far from trivial because of the diffraction limit [7]. As a result, arrays of non-interacting metallic nanostructures created on planar substrates have been usually employed for optical characterization [8]. When measuring such arrays, several elements are simultaneously illuminated at a time. As a result, the response of an isolated element is approximated by the averaged response of the illuminated structures, with the assumption that all of them are identical. This precludes to identify appropriately the performance of a single structure, with all the physical richness that it can provide. Scanning individual nanostructures requires strong focusing of the incident light [9,10]. This approach, however, besides being inefficient because light focusing is diffraction-limited and the nanostructures exhibit deep subwavelength cross-sections, is restricted to the measurement of a single element at a time, making unpractical the simultaneous measurement of multiple plasmonic elements built on a substrate. This limitation may become a roadblock in many applications, remarkably biosensing, which would ultimately require scanning multiple nanostructures in parallel and in real-time in order to get fully profit of the extreme miniaturization provided by plasmonics.

Exciting the metallic nanoparticles via dielectric waveguides is an appropriate way to solve this drawback: a single waveguide could both illuminate an individual nanostructure placed close to it and collect its response (or at least, part of it), enabling to measure its properties in a single shot. By using multiple waveguides simultaneously fed by a single optical source (for instance, using passive 1x$N$ power dividers), a number of metallic nanostructures could be simultaneously illuminated and their response measured in real time at the waveguides output. Notice that unlike metallic waveguides, dielectric waveguides can become practically lossless, enabling transfer of optical information between distant parts of a photonic chip at the centimeter scale and beyond. Owing to the ultrasmall dimensions of the metallic nanostructures under consideration, high-index waveguides enabling strong field concentration in transverse dimensions around $\lambda/2n$ (where $\lambda$ is the free space wavelength and $n$ is the waveguide core index) would be preferred. Amongst the available technological platforms for high-index integrated optics, silicon photonics presents several advantages, mainly an easy fabrication using standard microelectronics tools, ultimately enabling low-cost production at large-volumes [11-13]. In addition, whilst silicon can be used as a guiding element for near-infrared devices, silicon nitride waveguides enabling visible light guidance can be also produced in the same technological platform [14-15].

Recent experiments have demonstrated the excitation of subwavelength-size plasmonic nanostructures placed on top of silicon and silicon nitride waveguides at visible and infrared wavelengths [16-22]. The coupling between the guided field and the nanostructure takes place in the evanescent region of the waveguide, which is excited by the fundamental TE-like mode having the fundamental electric field component parallel to the chip plane. This results in relatively small interaction efficiencies (typically < 10% although ~19% is reported in [20]) so in many occasions a number of nanostructures need to be placed on the waveguide in order to produce observable effects (a dip in transmission) at the waveguide output [18,20]. Such weak coupling produces low power contrast (< 2) between maxima and minima in the measured spectra at the waveguide output, preventing its use in applications such as biosensing or all-optical switching which require high contrast to differentiate states. To overcome this issue and fully excite the nanostructure, it would be convenient to embed the nanostructure within a gap created in the waveguide, as we demonstrate numerically and experimentally in this work at telecom wavelengths.

## 2. Description of the concept

Our concept is schematically described in Fig. 1. A very small gap ($g$) separates two silicon waveguides (silicon nitride would provide a similar performance) with rectangular cross-section, being the plasmonic nanostructure placed in the middle of it and centered at the waveguide axis. This way, we ensure a maximum interaction of the propagating light field with the nanostructure, mainly in comparison with the approach in which the nanostructure is placed on top of the waveguide. As shown below, reflection at the gap boundaries can be relatively small. When illuminated, part of the light scattered by the nanostructure will be emitted towards the input port (backscattering) or towards the output port (forward scattering), interfering with the guided reflected and transmitted fields that have not interacted with the nanostructure, respectively. In addition, the metallic nanostructure will both scatter light out of the waveguides (out-of-plane but also in-plane) and absorb part of the incoming power, especially when a localized surface plasmon resonance (LSPR) is excited. Therefore, by measuring the optical power at both ports (transmission and reflection) we can get an accurate estimation of the absorbed and out-of-guide scattered power of the LSPR.

Notice the resemblance of our approach with that studied by R. Sapienza and colleagues [23] where a metallic nanoantenna is embedded within the silicon waveguide core, which is drilled by a hole with the nanoantenna dimensions. Although this approach enables a full excitation of the nanoantennas when perfectly aligned with the waveguide optical axis, it is limited in what refers to the fabrication (high-aspect ratio etching of silicon) as well as the type of metallic nanostructure to be inserted (in our approach, any kind of plasmonic nanostructure could be introduced, including elements with complex shapes supporting magnetic [24] or Fano resonances [25] or exhibiting directional scattering [26,27]).

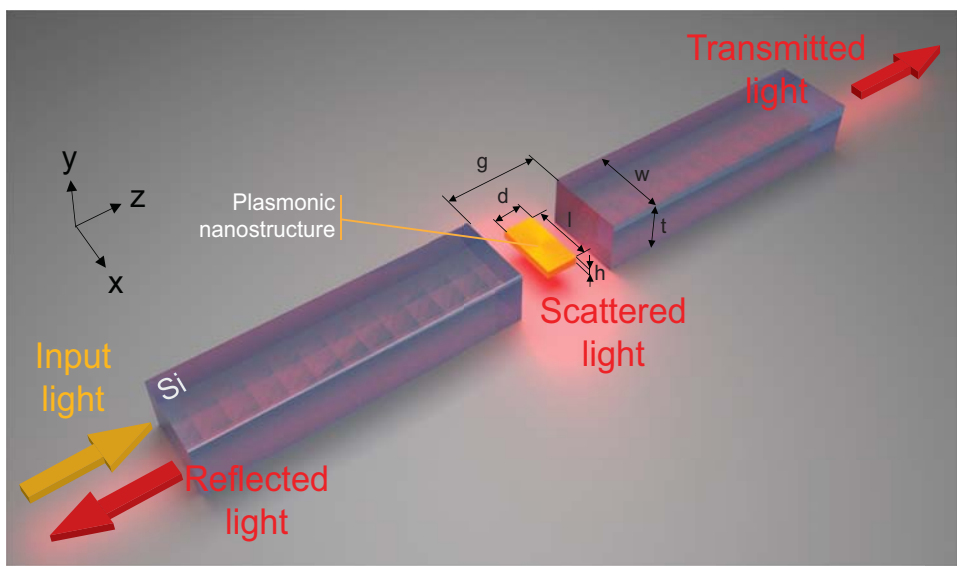

Fig. 1. Scheme of the proposed approach for full excitation of a plasmonic nanostructure (in this work as well as in the figure, we consider a gold nanostrip with dimensions $l{\times}d{\times}h$ placed inside a sub-micron gap (width $g$) created in a strip silicon waveguide with rectangular cross-section $w{\times}t$. The surrounding medium is silica, although other low-index dielectric (such as air) could be used.

## 3. Numerical results

Numerical simulations in this work have been performed using the commercial 3-D full-wave solver CST Microwave Studio, which implements finite integration technique. The structure was meshed using hexahedral mesh with 10 cells per wavelength, except near the metallic regions where the mesh has been refined up to approximately 15 nm (reaching around one hundred cells in total). Open boundary conditions (perfectly matched layers) are chosen for all external faces. We have considered that the strip waveguides, with dimensions $w$ = 400 nm and $t$ = 250 nm, are made of crystalline silicon and are surrounded by silica. The metallic element is made of gold, whose optical constants were obtained from ellipsometry measurements of thin films deposited using the same procedure as described below for the tested samples.

Before analyzing the complete structure we consider first the waveguide without any nanoparticle at the gap in order to know about the influence of the gap into the transmission and reflection behavior of the waveguide. Figure 2 shows the transmission $T$ and reflection $R$ spectra for the TE-like mode (a) and TM-like mode (b) for four different gap lengths in the wavelength window between 1.1 and 1.9 µm. To account for the waveguide spectral response, results in Fig. 2 are normalized with respect to the transmission spectra of the isolated waveguide (without gap). It is found that the shorter the gap is, the higher the transmission and the lower reflection will be, as could be expected by considering the light exiting the input waveguide will diverge because of the strong confinement and scattering will grow as the gap width increases. This also results in a reduction of the scattering losses for larger wavelengths. Notice that the gap can be considered to operate as a Fabry-Perot cavity with large radiative losses (light that is scattered out of the waveguides) and subwavelength length, which precludes the observation of ripples in the transmission spectra. We have also calculated the electric field profile in the middle of a 300 nm gap for both TE and TM-like modes. Results are shown in Figs. 2(c) and 2(d). It can be seen that, besides a large field component corresponding to the fundamental component of each guided mode ($E_x$ for the TE-like mode and $E_y$ for the TM-like mode) there is also a large longitudinal component $E_z$ with an asymmetric profile in both cases. This is related to the existence of a large transverse spin in silicon waveguides as a result of the strong confinement [28] and could be used for polarization manipulation when placing subwavelength nanoantennas in the proximity of the waveguide [29,30]. In our case, the existence of large longitudinal fields in the gap makes a strong difference in comparison with typical excitation of plasmonic nanostructures using free-space plane-wave-like light where the transverse component is dominant. Such longitudinal components must be carefully considered when trying to excite electric dipolar resonances (as, for example, in the case we are considering in this work) with the transversal field, mainly for large nanostructures or nanostructures placed out of the optical axis. But besides this, we can also consider the existence of such components as an interesting opportunity since it offers the possibility to play with the gap fields and the nanostructure shape and position to excite complex LSPRs that could results in effects not achievable with free-space excitations such as, for instance, directional scattering [26,27].

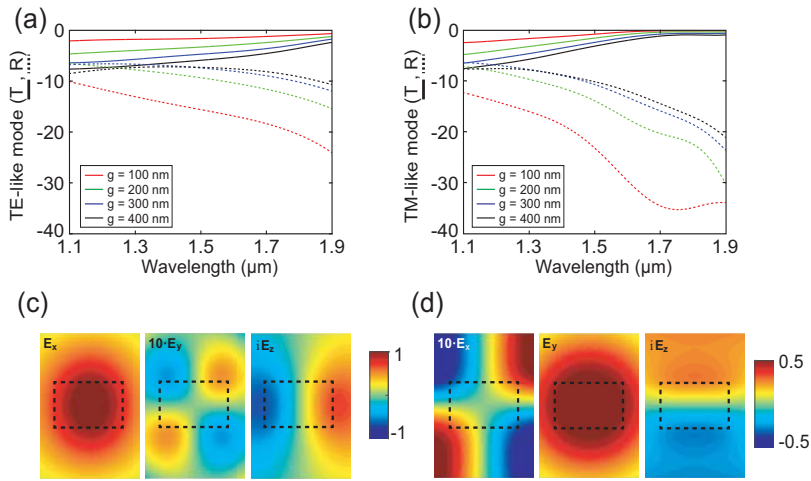

Fig. 2. Numerical study of the gap effects. Normalized transmission (T) and reflection (R) spectra for excitation using the fundamental TE-like (a) and TM-like (b) modes at different gap widths. The vertical axis is in dB units. Snapshot of the electric field components recorded at a plane placed at the middle of the gap ($g$ = 300 nm) for TE-like (c) and TM-like (d) waveguide modes.

After studying the response of the waveguide with gap, we consider now the response of the metallic element to be introduced in the gap when isolated. For simplicity, we choose a basic plasmonic nanostructure: a gold nanostrip with rectangular cross-section ($h$x$d$) and length $l$ whose dimensions can be chosen to exhibit an electric dipole LSPR at telecom wavelengths for excitation with an $x$-polarized incoming wave. Figure 3(a) shows the calculated scattering cross section (SCS) and absorption cross section (ACS) for a nanostrip with dimensions $l$ = 320nm, $d$ = 155 nm and $h$ = 40 nm surrounded by silica. The LSPR wavelength is ~1.75 μm though it can be tuned by modifying $l$. Note that the ACS value is multiplied by 5 to show more clearly that the maximum of the absorption is not located at the same wavelength as the maximum of the scattering, as expected. The permittivity retrieved from ellipsometric measurements of deposited thin films is also shown in Fig. 3(b), showing a good agreement with [31].

Finally, we consider the full system: the metallic nanostrip placed within the gap as shown in Fig. 1. To achieve a maximum interaction the nanostrip is aligned with the optical axis of the waveguide. In order to excite the LSPR, the TE-like mode of the silicon waveguide has to be launched so that a large $E_x$ component impinges on our nanostrip according to the results in Fig. 2(c). Notice that the longitudinal field component (see $E_z$ panel in Fig. 2(c)) will also have some effects on our metallic nanoparticle because of its length (320 nm), which does not take place in the SCS and ACS calculations where the nanoparticle is isolated and the incident field is purely $x$-polarized. When excited, the radiation emitted by the currents circulating on the nanostrip (here completely acting as a nanoantenna) will interfere with the transmitted and reflected guided waves, which will modify the responses shown in Fig. 2. Part of the emitted radiation will not be captured by the waveguides so will be lost, contributing to the scattering loss arising from the light coupling from waveguide to waveguide. Finally, there will be also a strong absorption, proportional to the ACS of the isolated nanostrip, as a result of the ohmic loss of the metallic nanostructure. Figure 3 also shows the transmission and reflection of the gold nanostrip embedded in a $g$ = 300 nm gap. Counterintuitively, there is not a minimum in transmission associated with the LSPR peak wavelength, as it happens when the metallic scatterer is embedded into the silicon waveguide [23]. Instead we observe that there is a crossing of the transmission and reflection curves in the region of the LSPR wavelength. Such a crossing can be considered as a signature of the excitation of the LSPR in our system. At wavelengths below the LSPR, we get large reflection and low transmission, being this situation reversed at wavelengths above the LSPR. This behavior can be explained by considering that the phase of the radiation emitted by the metallic nanostrip, which can be considered as a half-wavelength electric dipole antenna, changes with wavelength as expected for a polarizable particle, being the interference with the guided field constructive (destructive) for reflection (transmission) at wavelengths below the LSPR and vice versa for wavelengths above the LSPR. Remarkably, destructive interference in the transmitted signal reaches its maximum at wavelengths around 1.4 μm where the transmittance approaches zero. In this region, the interfering paths (the field transmitted from waveguide to waveguide and the field radiated by the nanostructure) will exhibit similar amplitudes but a $\pi$ phase shift between them. This confirms that the plasmonic nanostructure is fully excited by the incoming field, in contrast to the particle-on-top approach, for which the amplitude of the radiated field will be much smaller than that of the guided wave. In addition, we find that the reflection curve is in good agreement with the numerical results in [32] for a metallic nanoantenna placed at the output of a dielectric waveguide.

The asymmetry of the transmission spectrum profile can be well explained in the context of Fano resonances [33,34]. In this case, the optical signal propagating from waveguide to waveguide across the gaps plays the role of the broad continuum whilst the first-order LSPR of the gold nanostrip acts as discrete (although spectrally broad) resonance [35]. We fitted the transmission response to a Fano resonance with $F$ = 0.55, $\omega_0$ = 11.23·$10^{14}$ rad/s and $\gamma$ = 470 nm, where $F$ describes the degree of asymmetry, whilst $\omega_0$ and $\gamma$ correspond to the position and width of the isolated nanoparticle LSPR respectively [33]. The resulting curve is depicted in Fig. 3(a), showing a good agreement with the response obtained from numerical simulations.

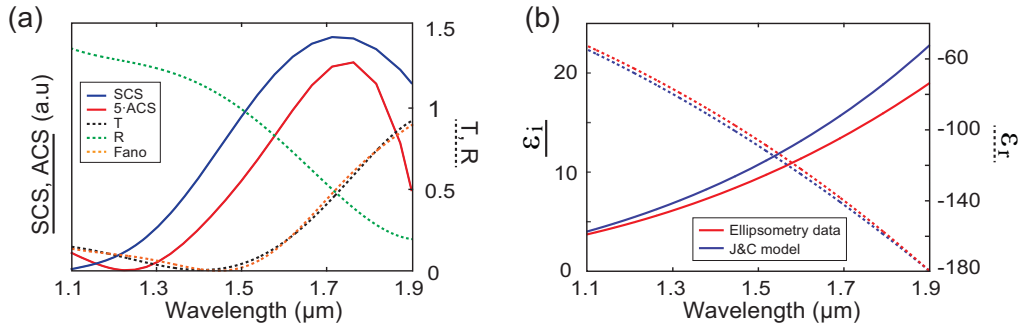

Fig. 3. (a) Numerical results when the gold nanostrip is included. Simulated scattering (RCS, blue solid), and absorption (ACS, red solid) cross-sections for a gold nanostrip with dimensions $l = 320$ nm, $d = 155$ nm and $h = 40$ nm, surrounded by silica. Transmission ($T$, black dashed) and reflection ($R$, green dashed) of the complet system with the nanostrip embedded in a $g = 300$ nm gap and Fano fitting (orange dashed) to the simulated transmission. (b) Electric permittivity of gold from ellipsometry of thin deposited layers (used in the numerical simualtions) and from the Johnson and Christy results [31].

It is possible to spectrally shift the transmission and reflection responses by changing the dimensions of the nanostrip. Results obtained when varying the dimensions $l$ and $d$ ($h$ is kept equal to 40 nm) for TE-like excitation are shown in Figs. 4(a) and 4(b), respectively. The curves are normalized with respect to the results without the nanostrip, this is, only considering the effects of the 300 nm gap. Regardless of the dimensions, we observe in all the cases the crossing point between the transmission and reflection curves. Remarkably, we observe that the normalized transmission is above 0 dB in the long wavelength region. This means that the scattering introduced by the gap is reduced when the metallic nanoantenna is inserted: a portion of the scattered field is captured by the nanoantenna and radiated towards the output waveguide, where it interferes constructively with the guided field producing an increase in the total transmission. When $l$ increases, we observe a red-shift of both the transmission dip and the transmission-reflection crossing point. This is a direct consequence of the nanostrip behaving as a half-wave dipole nanoantenna. In addition, we see that the transmission minimum is also modified when changing the nanoantenna dimensions. Indeed, we see in Fig. 4(b) that whilst a modification in $d$ does not produce appreciable changes in the transmission dip wavelength, it strongly modifies its depth, enabling values of the order of -50 dB, which is quite impressive if we consider that we have a scatterer with subwavelength dimensions. This available contrast in the transmission spectrum (>50 dB or five orders of magnitude) extremely exceeds the values achieved when illuminating metallic nanostructures placed on a substrate in an out-of-plane configuration or in the configuration when the nanostructure is placed on top of the waveguide, where the amplitude contrast in transmission (or reflection) is within one order of magnitude. The availability of such large contrast in the transmission levels may enable disruptive performance in several applications that make used of LSPRs of isolated nanostructure. For example, it would enable to get an ultralarge dynamic range for detecting substances if employed in biosensing. If utilized for switching purposes, it could allow for ON-OFF contrasts or modulation depths well above 10 dB, which should be sufficient to differentiate bits at the switch output.

Results for TM-like excitation are depicted in Figs. 2(c) and 2(d) for variations in $l$ and $d$ respectively. In this case, the LSPR is not excited because the electric field in the center of gap region has fundamentally a component along the y-axis and there are no appreciable features in the obtained spectra regardless of the nanostrip dimensions.

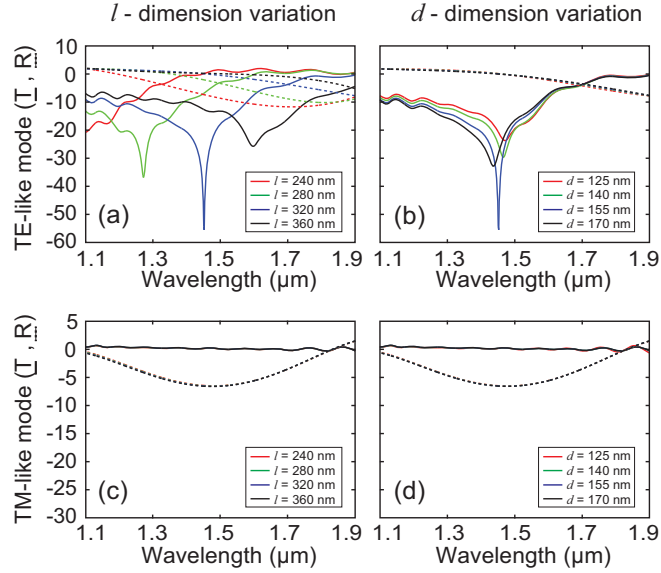

Fig. 4. Transmission and reflection through the silicon waveguide at TE-like (top) and TM-like (bottom) excitation for variations of the nanostrip dimensions ($l$ and $d$). Both transmission (solid line) and reflection (dashed line) are represented in dB.

## 4. Fabrication process

Here, we detail the fabrication process to embed the gold nanostrip in the silicon waveguide gap. The strip waveguides were fabricated on standard silicon on insulator (SOI) samples taken from SOITEC wafers with a top crystalline silicon layer thickness of 250 nm and a buried oxide layer thickness of 3 μm. The waveguides fabrication is based on an electron-beam direct-writing process performed on a coated 100 nm hydrogen silsesquioxane (HSQ) resist film. The mentioned electron-beam exposure, performed with a Raith150 tool, was optimized in order to reach the required dimensions employing an acceleration voltage of 30 KeV and an aperture size of 30 μm. After developing the HSQ resist using tetramethylammonium hydroxide, the resist patterns were transferred into the SOI samples employing an optimized inductively coupled plasma-reactive ion etching process with fluoride gases. After etching, a 105 nm-thickness silicon dioxide layer was deposited on the SOI sample by using a plasma enhanced chemical vapor deposition (PECVD) system from Applied Materials. This layer is deposited to center the metallic nanostructure with the optical axis of the waveguide, ensuring maximum interaction with the field within the gap. A second e-beam exposure prior to a metal evaporation and lift-off processes were carried out in order to define the 40 nm thickness gold nanoparticle inside the waveguide gaps. A 2 nm titanium layer were also evaporated to improve gold adhesion. Notice that this last step would allow for inserting metallic nanostructures with more complex shapes, such as split-ring resonators or bowtie nanoantennas. Finally, a micron-thickness silicon dioxide uppercladding was deposited on the SOI sample by using again PECVD in order to ensure homogeneity in the surroundings of the metallic nanostructure. Figure 5 shows scanning-electron-microscope (SEM) images of several fabricated samples. It can be seen that some deviations arisen in the fabricated samples, mainly in the position of the nanostructure within the gap. This could potentially produce different results when measuring the system in opposite directions. To account for this, we have characterized the fabricated structure in both transmission directions (from left to right and vice versa) and also for both guided modes (TE-like and TM-like), giving a set of four possible measurements. Figure 5 also describes the colour code that we have employed to describe the experimental measurements reported below.

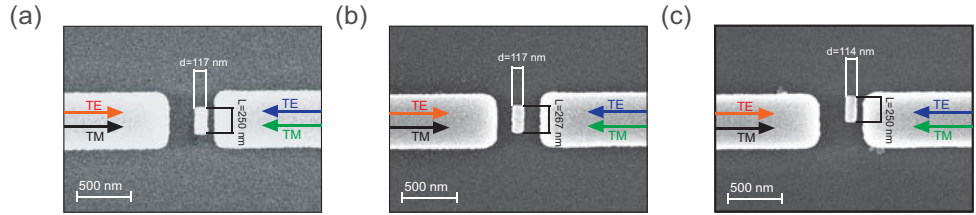

Fig. 5. SEM images of three fabricated structures. The top view of the fabricated systems is shown, with the top dimensions of the gold nanostrip (at the center) depicted in nm. The thickness of the nanostrip is 40 nm. Deviations in the nanoparticle position or dimensions were taken in account for the simulations.

## 5. Experimental results

The samples were characterized by using a standard butt-coupling system typically used to measure transmission spectra in photonic chips (see scheme in Fig. 6). As a light source, we used a tunable laser covering the range between 1260 and 1630 nm. A polarization controller was used to select the mode propagating along the silicon waveguides. A lensed fiber was used to couple light to the sample, a process in which we observed coupling losses of the order of 10 dB. At the output of the sample, we had an objective, a polarization filter and a splitter, which was used to split the light beam into two orthogonal paths: one directed to a near infrared camera to localize the output light spot and the other one directed to a power meter. The lensed fiber and the sample were placed on nanopositioners (not shown in the figure for the sake of clarity) which allow for aligning the input fibre with each waveguide included in the sample.

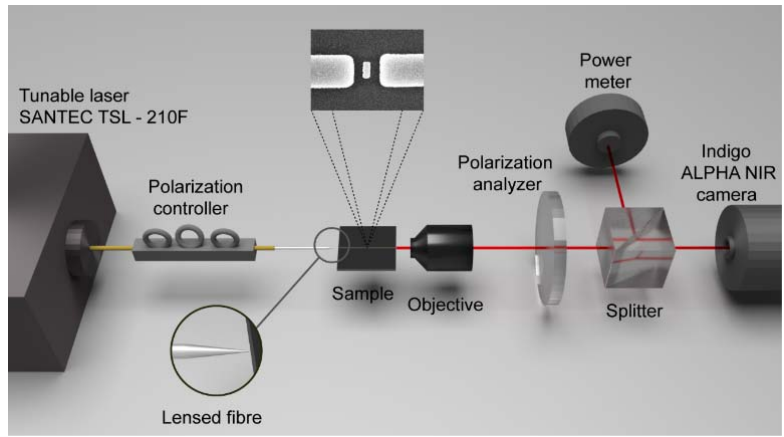

Fig. 6. Scheme of the optical set-up used in the experimental measurements. Insets show details of the lensed fiber and a SEM of the gap region of one of the tested samples.

In order to compare the transmission spectra obtained in simulations and measurements, we performed numerical calculations using the real geometrical data of the fabricated samples, including alignment deviations (see Fig. 5). The resulting transmission spectra once normalized with respect to the system without nanoparticle are depicted in Fig. 7. Since the range of wavelengths available for characterization is smaller than those shown in Fig. 3, we do not have access to all the details arising from the numerical simulations. Still, we can measure a relevant portion of the spectrum and, more importantly, we can observe the most interesting features of the structure for TE-like excitation: low transmission in the short wavelength region and transmission above 0 dB in the long wavelength region, with a 0 dB crossing close to the LSPR wavelength. For TM-like excitation, the transmission oscillates around 0 dB, as expected from the fact that the LSPR is not excited so the transmission with and without nanostrips remains almost the same. Notice that the experimental results match properly with the performed simulations. Remarkably, there are not huge difference between transmission results along counter-propagating paths even for the cases of significant misalignments. This can be explained by considering that the excitation of the LSPR is quite efficient in all the cases and the shift in the position of the nanotrips with respect to the ideal case are not sufficient to produce phase shifts that could result in observable changes in the transmission.

In Fig. 7 we see measured contrasts between transmission maxima and minima of the order of 10 dB. However, the fabricated structures show even larger contrast which are not shown in the transmission spectra because the minima fall out of the measurement window. Note that in the three cases the minima are out of the range of wavelength covered by the tunable laser. Numerical results give transmission minima of -11 dB at 1215 nm in case (a), -11 dB at 1155 nm in case (b) and -7.5 dB at 1270 nm in case (c). In all the cases, the large (> 10 dB) contrast are caused by the strong interaction between of the embedded nanonantenna and the waveguide field, which is not attainable in nanoparticle-on-top configurations, in which the nanostructure weakly interacts with the evanescent region of the waveguide mode.

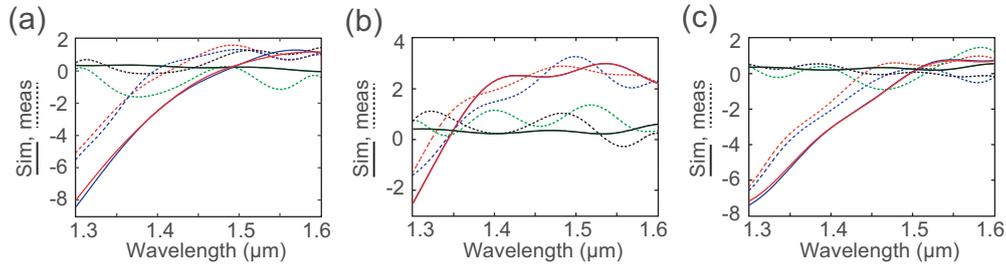

Fig. 7. Simulated and measured transmission ($T$, in dB) with respect to a waveguide with gap but without nanostrip. Experimental results are depicted in dashed line, while simulated results are depicted in solid line. TE-like and TM-like transmission from left to right (right to left) are depicted in red and black (blue and green), respectively. Each subfigure (a,b,c) corresponds to the structures shown in Fig 5 (a,b,c) respectively.

## 6. Conclusion

In summary, it has been proved that a subwavelength metallic nanostructure can be strongly excited by a guided mode by placing it at the gap of a discontinuous silicon waveguide. We have found that the observed resonance displays an asymmetric Fano behavior as a result of the interference between the guided mode and field radiated by the metallic nanoantenna. Remarkably, the nanostructure response is characterized by a crossing between the transmission and reflection spectra in the wavelength region close to the LSPR. The high contrast (> 10 dB in measurements, > 50 dB in numerical simulations) observed in transmission could be extremely helpful in applications including biosensing or switching.

**Acknowledgments**

We acknowledge support from the Spanish Ministry of Economy and Competiveness (MINECO) under grants TEC2014-51902-C2-1-R and TEC2014-61906-EXP the Valencian Conselleria d'Educació, Cultura i Esport under grant PROMETEOII/2014/034.